# Modulation of charge transport at grain boundaries in SrTiO$_3$: towards high thermoelectric power factor at room temperature

*Jianyun Cao,[†,‡,∥] Dursun Ekren,[†,§,∥] Yudong Peng,[†] Feridoon Azough,[†] Ian A. Kinloch,[†,‡,]* Robert Freer[†,]**

[†] Department of Materials, University of Manchester, Oxford Road, M13 9PL, U.K.

[‡] National Graphene Institute and Henry Royce Institute, University of Manchester, Oxford Road, M13 9PL, U.K.

[§] Department of Metallurgy and Materials Engineering, Iskenderun Technical University, 31200, Iskenderun, Hatay, Turkey

[∥] These authors contributed equally to this work.

ABSTRACT

Modulation of the grain boundary properties in thermoelectric materials that have thermally-activated electrical conductivity is crucial in order to achieve high performance at low temperatures. In this work, we show directly that the modulation of the potential barrier at the grain boundaries in perovskite $SrTiO_3$ changes the low temperature dependency of the bulk material's electrical conductivity. By sintering samples in a reducing environment of increasing strength, we produced $La_{0.08}Sr_{0.9}TiO_3$ (LSTO) ceramics that gradually change their electrical conductivity behaviour from thermally activated to single crystal-like, with only minor variations in Seebeck coefficient. Imaging of the surface potential by Kelvin probe force microscopy found lower potential barriers at the grain boundaries in the LSTO samples that had been processed in the more reducing environments. A theoretical model using band offset at the grain boundary to represent the potential barrier agreed well with the measured grain boundary potential dependency of conductivity. The present work showed an order of magnitude enhancement in electrical conductivity (from 85 to 1287 S cm$^{-1}$) and power factor (from 143 to 1745 µW m$^{-1}$ K$^{-2}$) at 330 K by this modulation of charge transport at grain boundaries. This significant reduction in the impact of grain boundaries on charge transport in $SrTiO_3$ provides an opportunity to achieve the ultimate "phonon glass electron crystal" by appropriate experimental design and processing.

**Introduction**

Grain boundaries exist in all polycrystalline materials and have significant impact on, and may dominate, the overall structural and functional properties of the material.[1-4] The modulation of grain boundaries' characteristics (e.g. grain size, local elementary composition and chemical state, etc.) provides an opportunity to tune bulk properties including mechanical strength,[5] electronic and ionic conductivity,[3, 6-8] photovoltaic efficiency,[4] and thermal conductivity.[9] Recently, interest has grown in the optimization and enhancement of thermoelectric properties of inorganic compounds via grain boundary engineering.[1, 3, 6-7]

Thermoelectric materials that can generate electricity efficiently from waste heat are of considerable interest in the development of future sustainable society.[10-13] Particularly, in the forthcoming age of internet of things (IoT) enabled by 5G network, thermoelectric generator that operates at room or near room temperatures can: (i) convert the low grade waste heat (30 to 250 °C) generated by electronic devices to electricity; (ii) power up wearable and mobile sensors using human body heat.[14-16] For thermoelectric materials, the energy conversion efficiency is evaluated by the dimensionless figure of merit ($ZT$), which is determined by equation: $ZT = (S^2\sigma/\kappa)T$, where $S$, $\sigma$, $\kappa$, and $T$ are Seebeck coefficient, electrical conductivity, thermal conductivity and absolute temperature, respectively; thus we need to maximise charge transport and minimise thermal transport.[11] The presence of grain boundaries affects both the power factor ($S^2\sigma$) and the thermal conductivity of bulk materials.[1, 9, 17] For some materials, such as $SrTiO_3$ and $Mg_3Sb_2$, the build-up of electrostatic potential at the grain boundary acts as a barrier for charge carrier transport, limiting the overall carrier mobility, particularly at and near room temperature.[18-21] Meanwhile, the grain boundary interface is itself a type of lattice defect assemblage that assists in the reduction of thermal conductivity.[9, 22] These conflicting effects of grain boundaries on electrical and thermal transport present major challenges in the development of high performance nano-crystalline thermoelectric materials operating over a range of temperatures. Therefore, modulation of grain boundaries, that can reduce their impact on charge carrier transport while retaining their ability to suppress lattice thermal conductivity, is in demand.

Recent work on n-type $SrTiO_3$-based thermoelectric ceramics showed that the thermally activated behaviour of low temperature electrical conductivity disappeared with use of certain processing conditions (temperature, oxygen partial pressure, etc.) and/or after incorporation of additives (e.g graphene).[3, 18, 22-23] The analysis of energy- and carrier concentration-independent weighted mobility showed that charge transport in polycrystalline $SrTiO_3$ can approach that of a single crystal.[7] Generally, this single crystal-like electrical conductivity at low temperature is attributed to the decrease of grain boundary resistance with the reduction of the built in potential barrier at the grain boundaries.[3, 6-7, 18] Localized changes in grain boundary chemistry and structure (elementary composition, chemical state, etc.) are thought to be the primary reasons for the reduction of the potential barrier at the grain boundaries.[3] However, to date, there has been no direct evidence, either experimental or theoretical, that links the grain boundary potential with the change in temperature dependency of conductivity in thermoelectric materials.

A recently developed two-phase model,[1] which treated the grain boundary as a separate phase with a band offset from the neutral grain phase, was able to capture the measured temperature dependent electrical conductivity in $Mg_3Sb_2$-based compounds. The band offset of the grain boundary phase acted as a potential barrier for carrier transport. This two-phase model showed good agreement with the grain size dependent conductivity in $Mg_3Sb_2$-based compounds,[1] as the contribution of grain boundary resistance becomes much less dominant at larger grain sizes. However, perovskite $SrTiO_3$-based ceramics, with comparably small grain sizes (0.5 to 5 µm) exhibited distinctly different temperature dependencies of conductivity, showing either thermally activated or single crystal-like behaviour.[22, 24] This inconsistency in charge transport in $SrTiO_3$-based ceramic with similar grain sizes strongly indicates that the intrinsic grain boundary properties (e.g. built-in potential) varies from sample to sample, even for samples with similar stoichiometry.

In this work, by combining experimental results with theoretical modelling, we directly show that the modulation of potential barrier at grain boundaries effectively switches the low temperature conductivity of La-doped $SrTiO_3$ from thermally activated to single crystal-like behaviour. We begin with the prediction of temperature dependent electrical conductivity using the recently developed two-phase model; by lowering the magnitude of band offset at the grain boundary (i.e. height of the potential barrier), the low temperature electrical conductivity changes from thermally activated to single crystal-like behaviour. We experimentally modulated the grain boundary properties of $La_{0.08}Sr_{0.9}TiO_3$ (LSTO) ceramic by control of processing condition. The as-prepared LSTO ceramic samples showed a gradual change of low temperature electrical conductivity from thermally activated to single crystal type, with an order of magnitude increase in electrical conductivity and power factor at 330 K. The experimental results fit well with the two-phase model, showing good agreement for the dependency of electrical conductivity on grain boundary potential. Moreover, the reduction in grain boundary potential was confirmed by imaging the surface and analysis of local potential by Kelvin probe force microscopy.

**Experimental section**

*Materials.* The starting powders of $TiO_2$ (> 99.9%) and $SrCO_3$ (> 99.9%) were obtained from Sigma Aldrich (Gillingham, Dorset, UK). $La_2O_3$ powder (> 99.99%) was obtained from PI-KEM (Magnus, Tamworth, UK). Isopropanol was supplied by Sigma Aldrich (Gillingham, Dorset, UK) and the graphene nanoplatelets, grade M25, were obtained from XG science (Lansing, USA).

*Preparation of $La_{0.08}Sr_{0.9}TiO_3$ (LSTO) powder.* The A-site deficient $La_{0.08}Sr_{0.9}TiO_3$ ceramic powder was prepared by a solid-state reaction approach. The starting powders of $TiO_2$, $SrCO_3$ and $La_2O_3$ were weighed according to the stoichiometry before mixing. The $La_2O_3$ powder was calcined in air at 1173 K for 6 h to remove moisture before weighing. After wet milling in isopropanol for 24 h using zirconia ball milling media, the well-mixed powders were dried at 90 °C for 24 h. The solid-state reaction to form LSTO was conducted in an alumina crucible at 1473 K in air for 8 h. The as-prepared LSTO powder was subjected to planetary milling (Retsch® Planetary Ball Mill PM 100) at 350 rpm for 4 hours to reduce the average particle size to ~ 590 nm.

*Sintering of LSTO ceramics.* The bulk LSTO ceramics were prepared by conventional pressureless sintering. A uniaxial press was used to compact the LSTO powders into green body pellets 15 or 20 mm in diameter and 5 mm in height. The as-formed pellets were densified at 1700 K in three types of conditions with increasingly strong reducing environment, namely, (1) $Ar-H_2-5\%$; (2) $Ar-H_2-5\%$ plus sacrificial carbon powder but not in direct contact with the LSTO green body; and (3) $Ar-H_2-5\%$ plus sacrificial carbon powder bed with the LSTO green body embedded; these three samples are labelled as $LSTO-H_2$, $LSTO-H_2-C$, $LSTO-H_2-in-C$, respectively. To create more oxygen vacancies, the sintering time (24 h) for the LSTO-H$_2$-C and LSTO-H$_2$-in-C samples is longer than that of LSTO-H$_2$ (12 h). The oxygen scavenging carbon powder bed is made of the aforementioned LSTO power + 5 wt.% graphene nanoplatelets. After sintering, the as-formed pellets were cooled in the corresponding environments to room temperature, and then cut into bars and discs of the appropriate sizes for detailed characterisation.

***Characterisation.*** The density ($\rho$) of the LSTO ceramic was determined using the Archimedes method. Crystal structure and lattice parameters were characterized by X-ray diffraction (XRD) using a Philips X'Pert diffractometer with Cu-K$\alpha$ source ($\lambda_{Cu-K\alpha}$ = 1.540598 Å). A continuous scan between 20° and 100° was recorded using 0.0167° step size and a dwell time of 6 s per step. X'Pert HighScore and TOPAS software were used for phase identification and Rietveld refinement. Grain size was undertaken on polished surfaces by scanning electron microscopy (SEM, TESCAN MIRA3 SC FEG-SEM), the linear intercept method was used to determine the average grain size.[25] X-ray photoelectron spectra (XPS) were collected with a Kratos Axis Ultra spectrometer using monochromatic Al K$\alpha$ radiation ($E_{source}$ =1486.69 eV). CasaXPS software was used for deconvolution of the Ti 2p core-level with a Shirley type background. Atomic force microscopy (AFM) and Kelvin probe force microscopy (KPFM) were performed using a JPK NanoWizard 4 XP NanoScience atomic microscope equipped with a Kelvin probe microscopy module. The images were recorded using a Pt-Ir coated silicon probe (SCM-PIT-V2, Bruker).

***Thermoelectric measurements.*** A ULVAC ZEM-3 system was used to simultaneously determine electrical conductivity and Seebeck coefficients; the measurements were performed at temperatures from 300 to 900 K in a low pressure helium atmosphere. This inert atmosphere with low oxygen partial pressure together with the medium high temperature limit of 900 K guarantee the stability of materials' property during measurements.[26-27] Thermal diffusivity ($D$) was determined using the laser flash method with a Netzsch LFA-457 laser flash apparatus in argon atmosphere. Differential Scanning Calorimetry (DSC, Netzsch DSC 404 F1 Pegasus) was used to measure the heat capacity ($C_p$); the measurements were performed in an argon atmosphere. The thermal conductivity ($\kappa$) was obtained from $\kappa = D\rho C_p$.

**Results and discussion**

**Figure 1a** to **d** illustrates the way that a negatively charged grain boundary forms and behaves as a potential barrier for electron transport in SrTiO$_3$ based materials. Based upon Snyder's work,[3, 7] the formation of potential barrier in n-type SrTiO$_3$ can be ascribed to the reduction of the concentration of the dominant point defect (oxygen vacancies) near the grain boundary (**Figure 1a**). Literature results from molecular dynamic simulation also indicate the depletion of oxygen vacancies in regions near dislocation cores,[28] which are analogue lattice defects to the grain boundaries. On the assumption that the concentrations of cations are constant across the grain boundary (**Figure 1b**), the depletion of positively charged oxygen vacancies (electron-donating defects) near grain boundaries compared to bulk grains induces difference in Fermi energy levels, promoting electron transfer from the grain to the grain boundary to maintain the equilibrium of Fermi energy level. This charge transfer leads to a negative potential at the grain boundary region, which alters the local electronic structure at the grain boundary by bending the conductive band upward (**Figure 1c**).[1, 3, 29] Energy band diagrams of two grains and their boundary region before and after contact are illustrated in **Figure S1a** and **b**, respectively. The trapped electrons at the grain boundary results in much less free carriers in the vicinity of grain boundary and thus greater resistance compared to the bulk neutral grain (**Figure 1d**). **Figure 1e** depicts the overall principal of a grain boundary potential in n-type La doped SrTiO$_3$ acting as a barrier for majority carrier (i.e. electrons) transport. In addition, the grain boundaries with lattice mismatch act as scattering centres for carrier transport. The combination of a potential barrier and lattice defects at the grain boundaries leads to compromised electrical conductivity compared to that of the bulk neutral grain. The two-phase model treats the grain boundary as a secondary phase with (i) a band offset ($\Delta E$) from the bulk grain phase (**Figure 1f**) representing the band bending, and (ii) a different carrier scattering mechanism from the bulk grain phase, mimicking carrier scattering, because of the lattice mismatch.

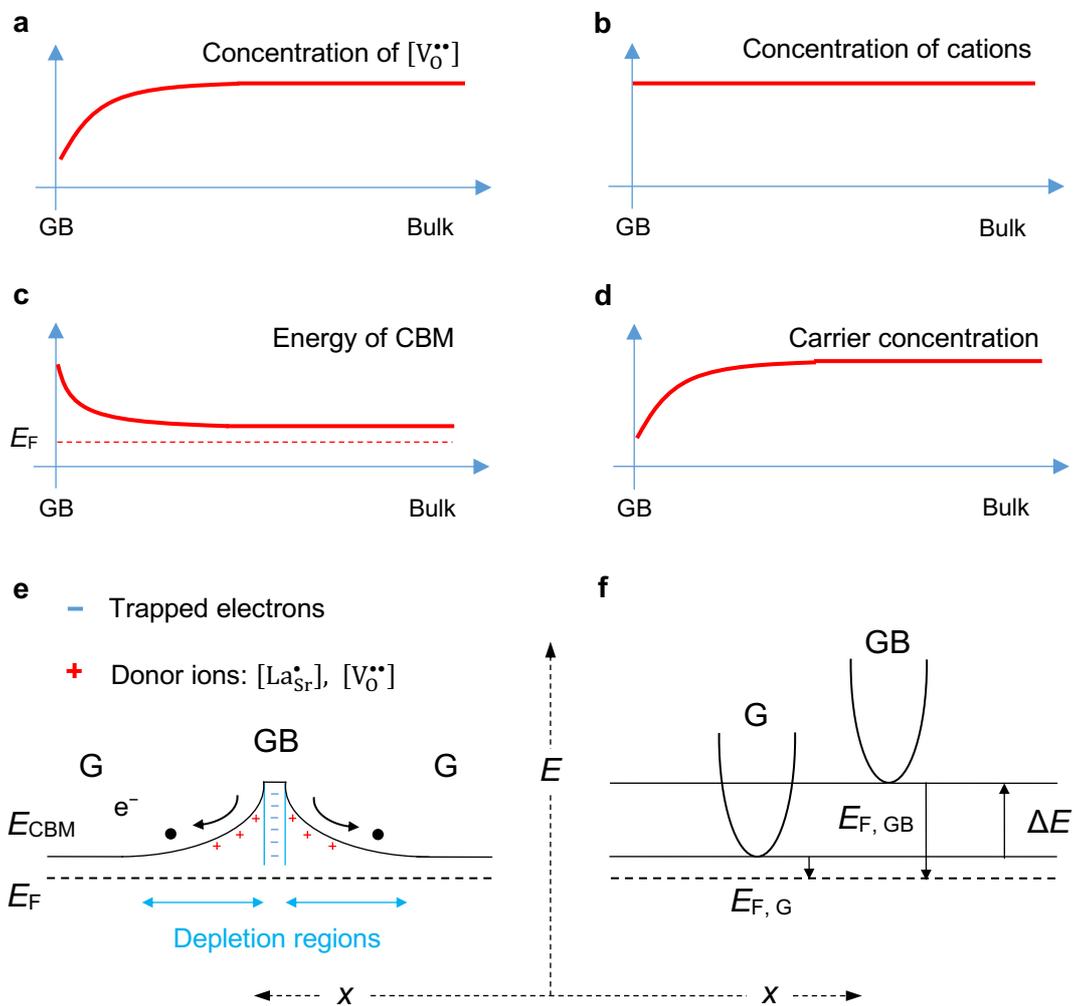

**Figure 1.** Diagrams showing that (a) oxygen vacancy is lower near grain boundary (GB) compared with the bulk grain, (b) concentrations of cations keep constant across grain boundary, (c) conductive band minimum (CBM) bends upward because of the equilibrium of Fermi energy level, (d) reduced concentration of free carrier at the vicinity of grain boundary.

(e) Schematic illustration of the potential barrier across the grain boundary, G represents a neutral grain, the black dots and arrows represent the transport of free electrons being inhibited by the grain boundary barrier. (f) Schematic illustration of the two-phase model for band offset (ΔE) in the grain boundary phase.

In detail, the Fermi energy level of a grain boundary phase ($E_{F,GB}$) measured from the band edge (CBM: conductive band minimum) is defined by the band offset to be:[1]

$$E_{F,GB} = E_{F,G} - \Delta E \tag{1}$$

where $E_{F,G}$ is the Fermi energy level of grain phase measured from the CBM, positive for free carrier energy and is higher than the CBM energy level.

Since the band offset $\Delta E$ also depends on the doping level of the material, an empirical, linear form of band offset function can be used to describe the energy dependency of $\Delta E$:

$$\Delta E = \Delta E_0 + a E_{F,G}$$

$$\tag{2}$$

where $\Delta E_0$ is the reference band offset at $E_{F,G} = 0$; it defines the magnitude of the band offset and therefore acts as a modelling parameter in this work; the coefficient $a$ is an empirical parameter which determines the energy dependency of $\Delta E$.

With a known Fermi energy level, the electrical conductivity ($\sigma$) and Seebeck coefficient ($S$) of the grain and grain boundary phases can be calculated by the well-established carrier transport equations:[30]

$$\sigma = \sigma_{E_0}(T) s F_{s-1}(\eta)$$

$$\tag{3}$$

$$S = \frac{k_B}{e} \left[ \frac{(s+1)F_s(\eta)}{s F_{s-1}(\eta)} - \eta \right]$$

(4)

where the transport coefficient $\sigma_{E_0}(T)$ is a temperature dependent, but energy independent coefficient that determines the magnitude of electrical conductivity; $T$ is absolute temperature, $s$ is the transport parameter related to the carrier scattering mechanism; $\eta$ is the reduced Fermi energy level ($\eta = E_F/k_BT$); $k_B$ is the Boltzmann constant; $e$ is the elementary charge, and $F_i(\eta)$ is the Fermi-Dirac integral

$$F_i(\eta) = \int_0^\infty \frac{\epsilon^i}{1+e^{\varepsilon-\eta}} d\epsilon$$

(5)

where $\epsilon$ is the reduced particle energy ($\epsilon = E/k_BT$; $E$: particle energy).

For the neutral grain phase of SrTiO$_3$ compounds, acoustic phonon scattering ($s = 1$) is assumed to be the dominant carrier scattering mechanism,[7, 22] with a crystalline metallic-type temperature dependency of transport coefficient ($\sigma_{E_0,G}(T) \propto 1/T$; the subscript G denotes grain phase). Earlier studies suggest that the donor concentration (point defects: [La$_{Sr}^{\bullet}$] or [V$_O^{\bullet\bullet}$]) dependency of carrier mobility in single crystal SrTiO$_3$ only becomes significant at low temperatures (e.g. < 150 K).[17, 31] The present work is focusing on room temperature and above and thus the ionic impurity scattering is not included as a scattering source in bulk neutral grain. We verified this assumption by fitting literature data for carrier transport for La or Nb doped SrTiO$_3$ single crystals,[32] using the value of $\sigma_{E_0,G}$ at 300 K as a fitting parameter. The modelled results with $\sigma_{E_0,G}$ = 900 S cm$^{-1}$ at 300 K (**Figure S2**) agree well with the reported temperature dependency of electrical conductivity, as well as the log |S|−log $\sigma$ plot. Hence, $s = 1$ and $\sigma_{E_0,G}(T) \propto 1/T$ with a value of 900 S cm$^{-1}$ at 300 K were used as universal modelling parameters for the neutral grain phase in this work. For the grain boundary phase with space charge and lattice mismatch, an ionized-impurity scattering model ($s = 3$;

$\sigma_{E_{0,GB}}(T) \propto T^3$, the subscript GB denote grain boundary phase) was used,[1] with $\sigma_{E_{0,GB}}= 0.15$ S cm$^{-1}$ at 300 K. **Table 1** shows the optimized modelling parameters for our materials.

A simple series circuit model was used to calculate the overall charge transport behaviour; the overall electrical conductivity $\sigma$ was obtained from:

$$\frac{1}{\sigma} = \frac{1-t_{GB}}{\sigma_G} + \frac{t_{GB}}{\sigma_{GB}}$$

(6)

where $t_{GB}$ is the size fraction of grain boundary phase in the ceramic. The value of $t_{GB}$ is proportional to the grain size thus can be estimated from microstructure characterization. This estimation of $t_{GB}$ from grain size assumes that the thickness of the carrier-depleted region is consistent for different samples, i.e. the thickness of the carrier depletion region is independent of the magnitude of the potential barrier. This assumption is reasonable as the SrTiO$_3$-based ceramics developed for thermoelectric applications have high donor concentrations, typically ~10$^{20}$ cm$^{-3}$, therefore the charge screening length should be similar. Although a more accurate description of a real grain boundary should correlate the thickness of the depletion region to the magnitude of potential barrier,[33] we found the current approximation is sufficiently good to fit the experimental results.

**Table 1**. Modelling parameters for La$_{0.08}$Sr$_{0.9}$TiO$_{3-\delta}$ (LSTO) prepared with different sintering conditions.

| Sample | $t_{GB}$ [a] | Band offset function | | Transport coefficient (300 K) | |
| --- | --- | --- | --- | --- | --- |
| | | a | $\Delta E_0$ | $\sigma_{E_{0,G}}$ | $\sigma_{E_{0,GB}}$ |
| LSTO H$_2$ | 0.001 (1.75 μm) | | 100 meV | | |
| LSTO H$_2$-C | 0.0005 (3.69 μm) | 0.3 | 75 meV | 900 S cm$^{-1}$ | 0.15 S cm$^{-1}$ |
| LSTO H$_2$-in-C | 0.0005 (2.67 μm) | | 10 meV | | |

[a] the average grain size of the sample is shown in the parenthesis

For the overall Seebeck coefficient $S$, in view of the small fraction of grain boundary phase in the ceramic ($t_{GB} < 0.001$), we followed the recent work of Kuo et al.,[1] and assumed

$$S \approx S_G \qquad (7)$$

where $S_G$ is the Seebeck coefficient of the bulk grain; see detailed explanations in supporting information and **Figure S3**. Note that the energy filtering effect of grain boundary barriers on Seebeck coefficient only becomes prominent for nanostructured materials with grain sizes < 50 nm or size fraction of barrier phase being high.[34-35] The present work focuses on $SrTiO_3$ based ceramics that have grain sizes over 1 μm and thus their Seebeck coefficients should be dominated by carrier concentration. The assumption of $S \approx S_G$ allowed extraction of the reduced Fermi energy level $\eta$ of the grain phase, using equation (4), from the experimentally measured overall Seebeck coefficient of the ceramic. In turn, this allows calculation of the reduced Fermi energy level of the grain boundary phase using the band offset function (1). Subsequently, the carrier transport characteristic of the polycrystalline $SrTiO_3$ was modelled with only one variable, $\Delta E_0$.

To demonstrate the approach, we modelled the impact of $\Delta E_0$ on the temperature dependencies of electrical conductivity (**Figure 2**a); the set of Seebeck coefficients for determining the reduced Fermi energy level of the grain phase was obtained from the LSTO-$H_2$ sample that will be discussed later. Obviously, with the decrease of $\Delta E_0$ from 80 to 10 meV, the temperature dependency of conductivity at low temperatures (300 to 450 K) switches from a thermally activated type to a metallic type. The dramatically enhanced conductivity suggests high power factor values, as the increase of conductivity is due to the change of grain boundary property rather than a change of the carrier concentration of the neutral grain phase (i.e. Seebeck coefficients remain unchanged). The plots of modelled log $|S|$–log $\sigma$ and power factor–log $\sigma$ (**Figure 2b**) clearly illustrate the enhancement in the powder factor by reduction of grain boundary barrier heights. For instance, with a Seebeck coefficient of 100 μV K$^{-1}$, the reduction of $\Delta E_0$ from 80 to 10 meV leads to the enhancement of power factor from 360 to 1560 μW m$^{-1}$ K$^{-2}$ (labelled in **Figure 2b**).

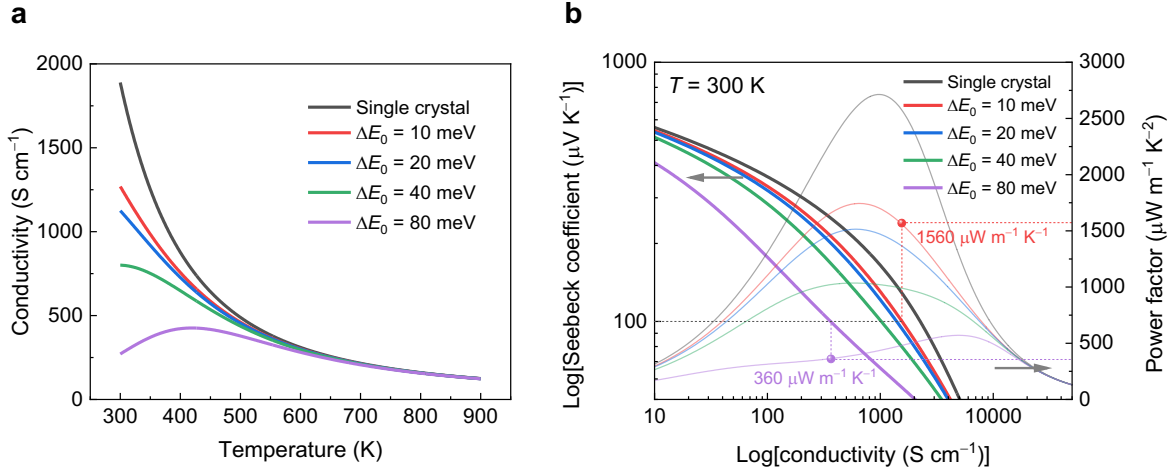

**Figure 2.** Theoretical predictions for grain boundary (GB) dominated charge transport in SrTiO$_3$ ceramics using the two-phase model. (a) Modelled temperature dependency of electrical conductivity for SrTiO$_3$ ceramics with different reference band offset values ($\Delta E_0$) and a fixed $t_{GB}$ value of 0.0005; other parameters ($a$, $\sigma_{E_0,G}$ and $\sigma_{E_0,GB}$) are kept at the universal values shown in Table 1. (b) Simulated log $|S|$—log $\sigma$ plots (thick solid lines) and plots of power factor as a function of log $\sigma$ (feint lines) at room temperature (300 K) for SrTiO$_3$ ceramics with different values of $\Delta E_0$. The modelled results for a single crystal ($t_{GB}$ = 0; solid black line) are added for reference purposes.

We then fitted literature data for electrical conductivity for SrTiO$_3$-based ceramics,[18, 36] using the two-phase model. Typical fitting results are presented in **Figure 3**. We found a set of universal fitting parameters ($a$ = 0.3, $\Delta E_0$ = 100 meV, $t_{GB}$ = 0.00025, $\sigma_{E_0,G}$ = 900 S cm$^{-1}$, $\sigma_{E_0,GB}$ = 0.15 S cm$^{-1}$) is sufficient to provide good agreement with the experimental results. In addition, for La$_x$Sr$_{1-y}$TiO$_3$ with medium high doping level ($x$ = 0.125 to 0.15), the high electrical conductivity due to high carrier concentration overwhelmed grain boundary resistance at low temperatures, leading to a metallic-type conductivity. However, this way of achieving high conductivity at low temperatures, by heavy doping, has the disadvantage that Seebeck coefficients are severely reduced. In fact, a small Seebeck coefficient, down to ~65 µV K$^{-1}$ at 470 K as shown in **Figure 3a and c**, is required to give the neutral grain sufficiently high electrical conductivity to overwhelm the grain boundary resistance at room temperature (**Figure 3b and d**).

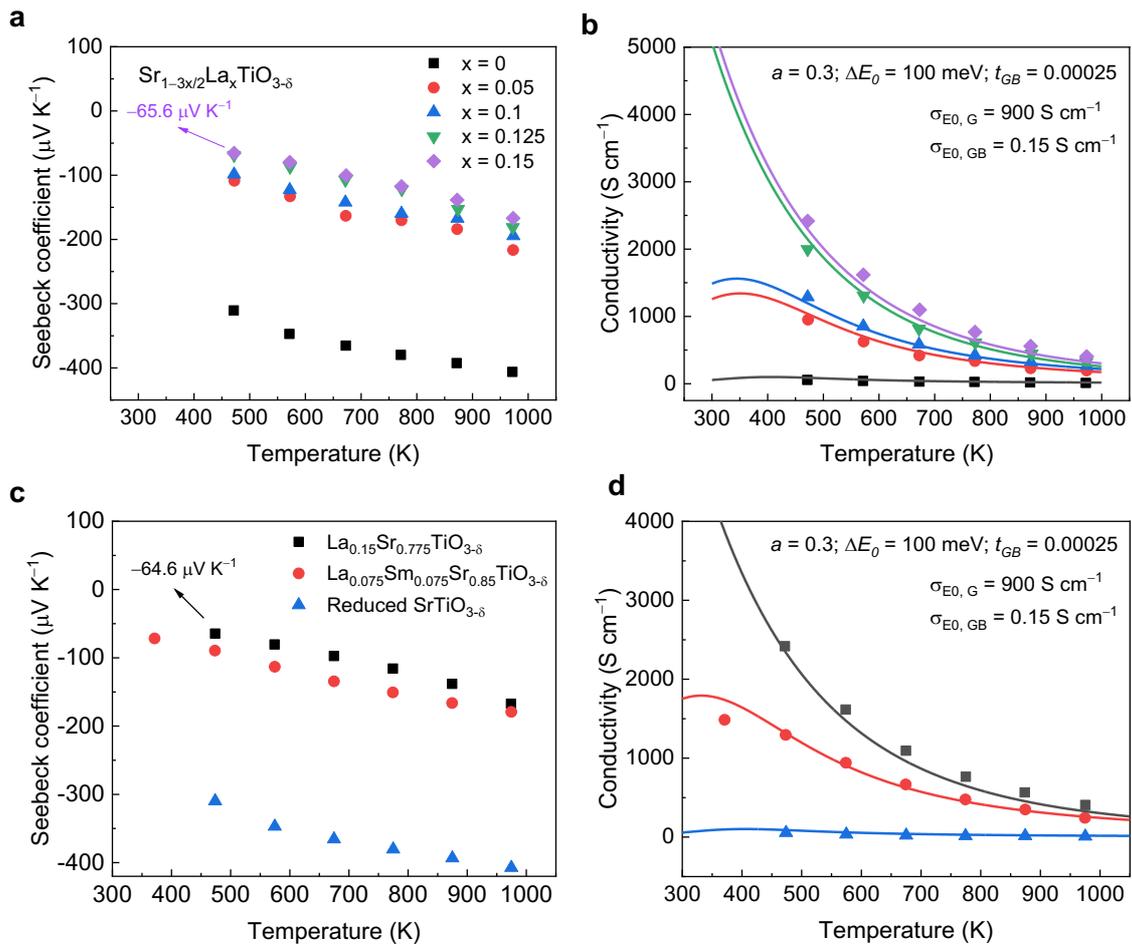

**Figure 3**. Fitting of the measured carrier transport results from the literature using the two-phase model. Experimental data points are displayed as symbols and the modelled results are shown as solid lines. Data points for Seebeck coefficients (a) and electrical conductivity (b) are from the work of Lu et. al.[36] Data points for Seebeck coefficient (c) and electrical conductivity (d) are from Boston et. al.[18]

The current literature lacks results from a set of pure SrTiO$_3$-based samples that have the same stoichiometry and comparable Seebeck coefficients, but distinctly different electrical conductivity, to allow a detailed study of grain boundary effects. Nevertheless, an earlier report,[17] together with recent investigations,[3, 18, 22] of SrTiO$_3$-based electronic ceramics imply it should be possible to experimentally modulate the grain boundary property by the control of processing conditions (temperature, oxygen partial pressure, etc.). The change of processing conditions, such as higher temperature and/or more reducing atmosphere with lower oxygen partial pressure, lead to a transformation of low temperature conductivity from thermally activated to a metallic type.[18, 37] This change of conductivity behaviour is suspected to be predominantly due to the modulation of grain boundaries (e.g. a localized increase in concentration of oxygen vacancies).[17]

Therefore, we experimentally modulated the grain boundary properties (e.g. oxygen vacancy concentration) of La$_{0.08}$Sr$_{0.9}$TiO$_3$ (LSTO) ceramic via control of the reducing strength of the sintering environment (indirect control of oxygen partial pressure). Sample preparation details are illustrated in **Figure 4** and described in the experimental section. The relative densities for the sintered LSTO-H$_2$, LSTO-H$_2$-C and LSTO-H$_2$-in C samples are 94.0%, 96.0% and 89.3%, respectively.

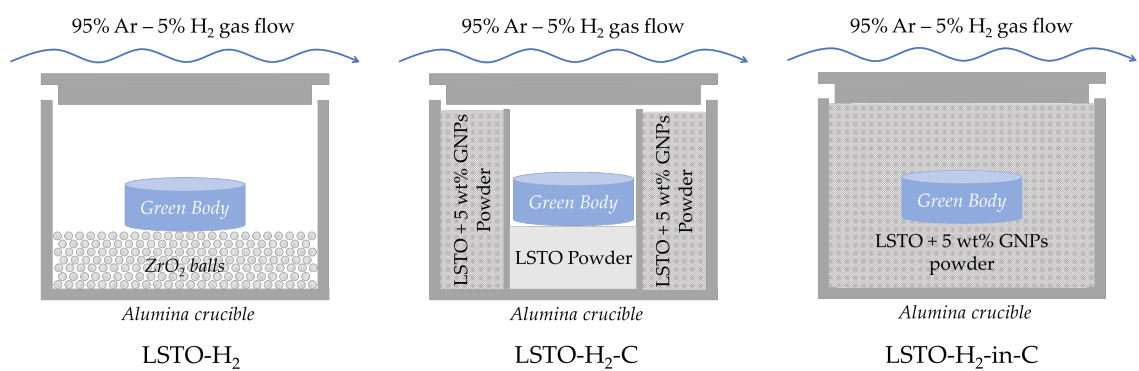

**Figure 4.** Schematic illustrations for the sintering of LSTO ceramics under different conditions with increasingly strong reducing environment from left to right. The oxygen scavenging sacrificial carbon powder bed is made of LSTO + 5 wt% graphene nanoplatelets (GNPs).

X-ray diffraction (XRD) patterns collected from the three LSTO samples (**Figure 5a**) show a cubic perovskite structure with *Pm-3m* space group symmetry. The decrease of diffraction angle for the samples prepared in more reducing environments (inset of **Figure 5a**) suggests an increase in the lattice parameter (**Table 2**). This increase corresponds to an increasing fraction of the small sized $Ti^{4+}$ (60.5 pm) being converted to relatively larger sized $Ti^{3+}$ (67 pm).[23] X-ray photoelectron spectroscopy (XPS) further confirmed the increasing degree of reduction of the LSTO samples from the more reducing environments. The deconvolution of Ti 2p core-levels (**Figure 5b**) leads to quantization of the concentration of $Ti^{3+}$ as a fraction of total Ti, i.e. ($Ti^{3+}$/Ti); the $Ti^{3+}$ concentrations for the LSTO-$H_2$, LSTO-$H_2$-C and LSTO-$H_2$-in-C are 3.0%, 3.2% and 4.4%, respectively. This enhancement in the $Ti^{3+}$ concentration implies a higher concentration of oxygen vacancies according to

$$[Ti'_{Ti}] \approx [La^{\bullet}_{Sr}] + 2[V^{\bullet\bullet}_O] - 2[V''_{Sr}] \tag{8}$$

The determination of $Ti^{3+}$ concentration also allows the estimation of carrier concentration (*n*) by[38]

$$n = (Ti^{3+}/Ti) \times N_{fu}/V_{uc} \tag{9}$$

where $N_{fu}$ is the number of formula per unit cell (1 for $SrTiO_3$) and $V_{uc}$ is the volume of unit cell. The calculated results are displayed in **Table 2**. The LSTO-$H_2$-in-C shows the highest carrier concentration of $7.36 \times 10^{20}$ $cm^{-3}$.

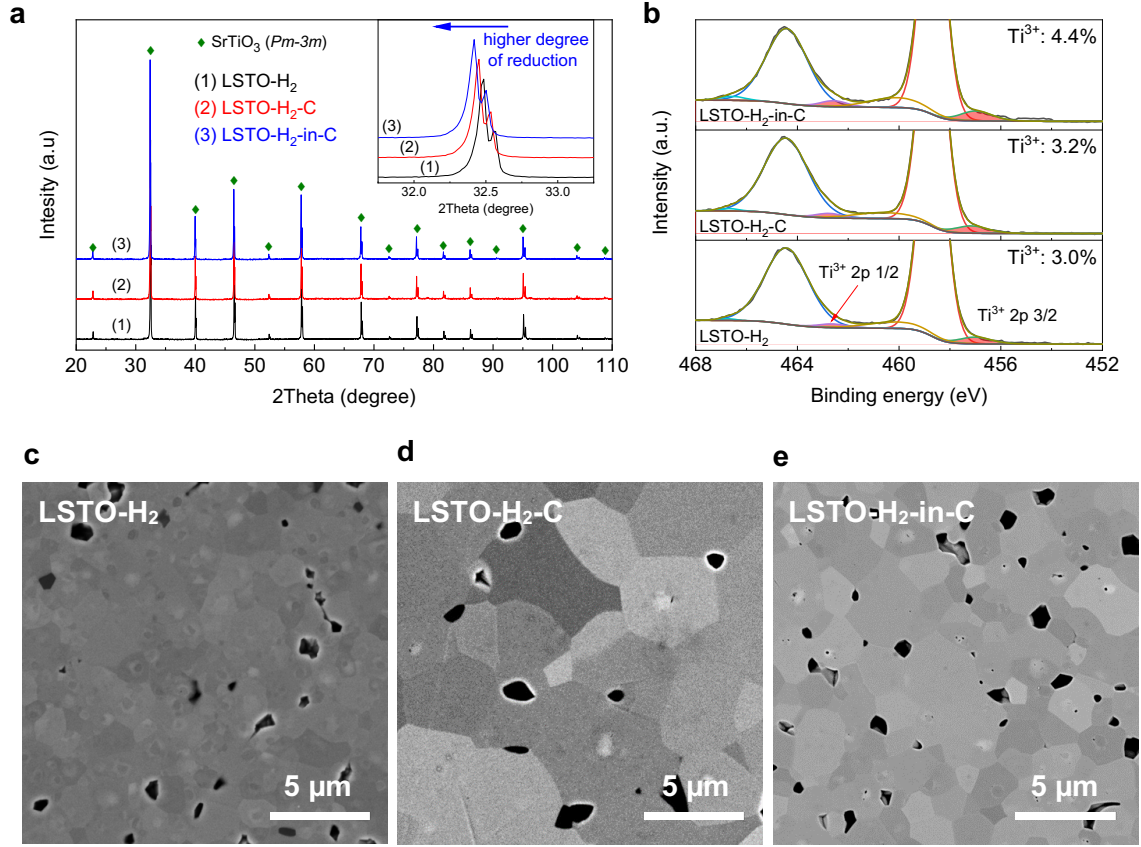

**Figure 5**. Characterization of the LSTO ceramics sintered at three different conditions with increasingly strong reducing environment. (a) XRD pattern for the three LSTO samples. (b) XPS Ti 2p core levels for the three LSTO samples. (c), (d) and (e) SEM micrographs showing grain sizes of the LSTO-H$_2$, LSTO-H$_2$-C and LSTO-H$_2$-in-C samples, respectively.

**Table 2**. Carrier concentration estimated from the Ti$^{3+}$/Ti, and the corresponding drift mobility and effective mass at 330 K.

| Sample | $a$ (Å) | Ti$^{3+}$/Ti (%) | $n$ ($10^{20}$ cm$^{-3}$) | $\sigma$ (S cm$^{-1}$) | $S$ (μV K$^{-1}$) | $\mu$ (cm$^2$ V$^{-1}$ s$^{-1}$) | $m^*/m_0$ [a] |
|---|---|---|---|---|---|---|---|
| LSTO-H$_2$ | 3.9080 | 3.0 | 5.02 | 85 ± 4 | −129 ± 9 | 1.1 | 3.7 |
| LSTO-H$_2$-C | 3.9087 | 3.2 | 5.39 | 492 ± 25 | −116 ± 8 | 5.7 | 3.5 |
| LSTO-H$_2$-in-C | 3.9095 | 4.4 | 7.36 | 1288 ± 64 | −115 ± 8 | 10.9 | 4.2 |

[a] free electron mass ($m_0$)

**Figure 6** displays the temperature dependent carrier transport properties for the three LSTO ceramic samples. The absolute value of Seebeck coefficient (**Figure 6a**) is slightly lower for the samples sintered in the more reducing environments, corresponding to the higher carrier concentrations implied by the $Ti^{3+}$ concentration (**Table 2**). In contrast to the relatively small variation in Seebeck coefficient, the three LSTO samples exhibit distinctly different electrical conductivities, particularly at temperatures below 500 K (**Figure 6b**). For example, at 330 K there is over one order of magnitude difference in electrical conductivity between the LSTO-$H_2$ (85 S cm$^{-1}$) and the LSTO-$H_2$-in-C (1287 S cm$^{-1}$) samples. Considering the variation of density among these three samples, we estimated their effective conductivities at 330 K following literature reported approach based on Maxwell equations,[39] see details in Supporting Information and Table S2. The results show that the effective electrical conductivity is higher than the experimental values, and the conductivity increase due to correction is more pronounced for the samples containing higher degree of porosity.

In addition to distinct differences in the magnitude of electrical conductivity, the as-prepared samples also show very different temperature dependencies, particularly at low temperature. Both of the LSTO-$H_2$ and LSTO-$H_2$-C samples show a typical grain boundary dominated, thermally-activated conductivity, i.e., at low temperatures the electrical conductivity increases, reaches a maximum and then decreases at higher temperatures. Due to the larger grain size of LSTO-$H_2$-C, and plausibly lower grain boundary potential barrier compared to LSTO-$H_2$, the LSTO-$H_2$-C has higher electrical conductivity at low temperatures and peak electrical conductivity at a lower temperature (smaller thermal energy needed to overcome the grain boundary resistance). More interestingly, although LSTO-$H_2$-in-C has a smaller grain size (2.67 μm) than that of the LSTO-$H_2$-C (3.69 μm), the LSTO-$H_2$-in-C sample exhibits a metallic type (i.e. single crystal-like) electrical conductivity in the measured temperature range from 300 to 900 K, with the highest conductivity up to 1423 S cm$^{-1}$ at

~300 K. This switch of temperature dependency is primarily due to the change of the grain boundaries' property (barrier height) rather than the grain size.

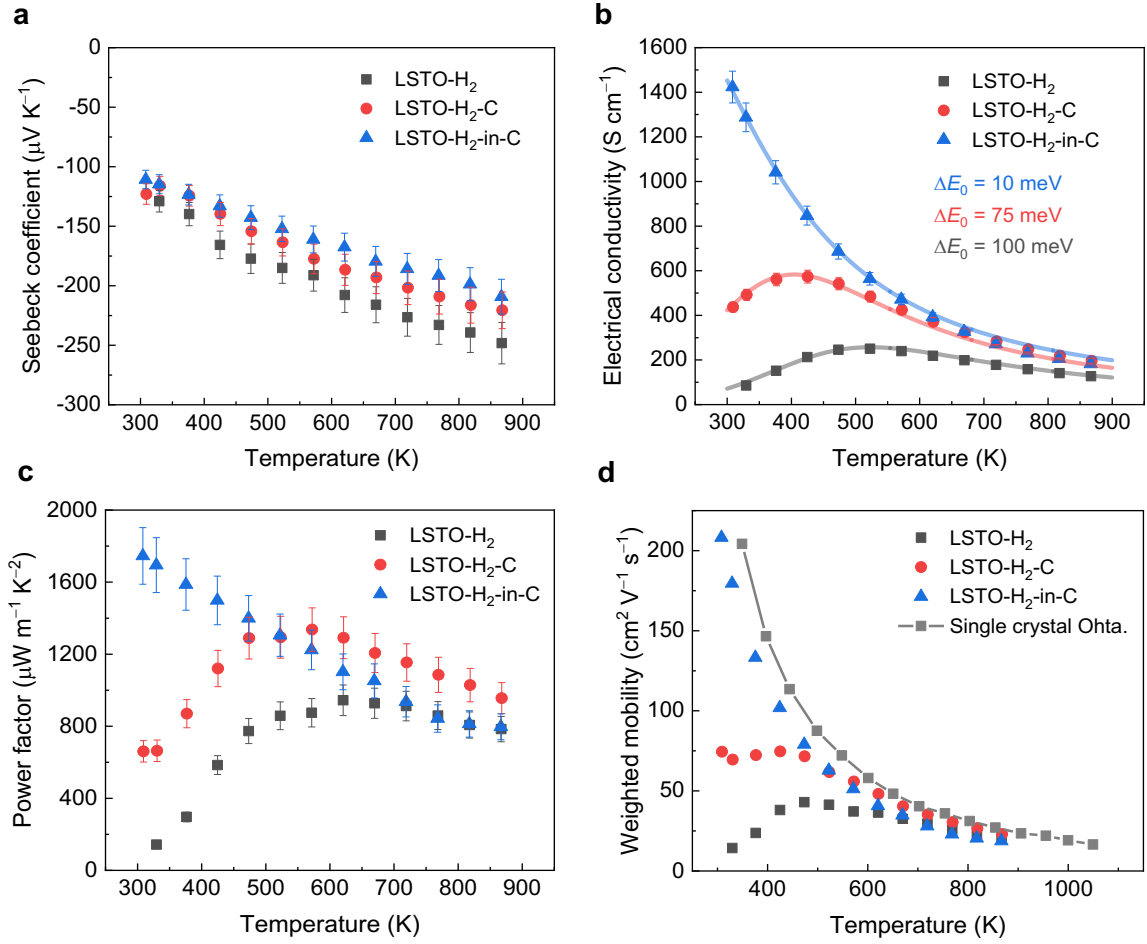

**Figure 6**. Temperature dependent electric charge transport properties of as-prepared LSTO ceramic samples; experimental data are represented by solid symbols. (a) Seebeck coefficient, (b) electrical conductivity, solid lines are from the model, using the parameters listed in Table 1; the simulated curves well capture the evolution of electrical conductivity with temperature. (c) Power factor, (d) weighted mobility; the weighted mobility of single crystal La doped $SrTiO_3$ calculated from the literature data[32] is included for comparison purposes.

To provide further insight into the grain boundary processes, we firstly calculated the electrical conductivity of neutral grain phase using equation (3) and the reduced Fermi energy level $\eta$ extracted from the measured Seebeck coefficients (**Figure 6a**) according to equation (4). The calculated conductivities of the neutral grain phase for the LSTO-$H_2$, LSTO-$H_2$-C and LSTO-H$_2$-in-C samples are quite similar, particularly the latter two (**Figure S4**). This decoupling of the neutral grain phase electrical conductivity from the measured overall conductivity directly indicates the dominance of grain boundaries in carrier transport in polycrystalline SrTiO$_3$-based ceramics. We subsequently fitted the measured electrical conductivity data with the two-phase model (equation (6)); using parameters listed in **Table 1**; the reduced Fermi energy level of the grain boundary phase was calculated using equation (1) and band offset function (equation (2)), with $\Delta E_0$ as the only fitting parameter. The value of $t_{GB}$ was chosen according to average grain size from SEM images (**Figure 5**c, d and e) and empirically according to the literature[1] as discussed above, and is therefore considered as a "known" parameter. The solid lines displayed in **Figure 6**b are from the two-phase model; the simulated results fit well to the measured electrical conductivity. The $\Delta E_0$ values that lead to the best fit of the measured conductivity data are 100, 75 and 10 meV for LSTO-H$_2$, LSTO-H$_2$-C and LSTO-H$_2$-in-C samples, respectively.

The power factor ($S^2\sigma$) exhibits similar temperature dependency as that of the electrical conductivity (**Figure 6**c). The LSTO-H$_2$-in-C has the highest power factor value of 1745 µW m$^{-1}$ K$^{-2}$ at room temperature, an order of magnitude higher than that of the LSTO-H$_2$ sample (143 µW m$^{-1}$ K$^{-2}$) at 330 K. In addition, the LSTO-H$_2$-C also shows a high power factor value of 1337 µW m$^{-1}$ K$^{-2}$ at 570 K. These power factor values are among the highest values reported for SrTiO$_3$ based compounds (**Table S1**).[6, 18, 20-24, 36, 38, 40-44]

We further computed the weighted mobility ($\mu_w$) of the LSTO samples from the measured electrical conductivity and Seebeck coefficients using the approach reported recently by Snyder et.al.[7, 45] In detail, the LSTO ceramic was treated as a homogeneous phase with acoustic phonon scattering as the dominating scattering mechanism (transport parameter $s =$ 1), the reduced Fermi energy level $\eta$ at each temperature was extracted from the Seebeck coefficients using equation (4), and then the temperature dependent transport coefficient $\sigma_{E_0}(T)$ was computed according to equation (3). The determination of transport coefficient $\sigma_{E_0}(T)$ then leads to the calculation of weighted mobility by:

$$\sigma_{E_0} = \frac{8\pi e(2m_e k_B T)^{3/2}}{3h^3}\mu_w \quad (10)$$

where $m_e$ is the free electron mass, $h$ is the Plank constant. This weighted mobility is temperature dependent but energy and carrier concentration independent, and allows a direct comparison of carrier mobility for samples at various doping levels. The comparison of weighted mobility derived from the literature data for SrTiO$_3$-based single crystals[32] shows similar values of weighted mobility, regardless of the doping concentration (**Figure S5**). **Figure 6d** shows the computed $\mu_w$ for the three LSTO samples prepared in the present work. Significant enhancement of $\mu_w$ is obvious for the LSTO prepared in the most reducing environment; it has the smallest $\Delta E_0$ value from the two-phase model. Excitingly, the LSTO-H$_2$-in-C sample with a metallic type electrical conductivity at the measured temperature range exhibits a weighted mobility approaching that of the single crystal (**Figure 6d**).

The carrier concentration $n$ determined from compositional data by **equation (9)** also allows the estimation of drift mobility ($\mu$) from the measured electrical conductivity using the Drude–Sommerfeld free electron model

$$\sigma = ne\mu \quad (11)$$

Further, with the parabolic band and energy-independent scattering approximation, the carrier effective mass ($m^*$) was estimated from the measured Seebeck coefficients by[38]

$$S = \frac{8\pi^2 k_B^2}{3eh^2}m^* T \left(\frac{\pi}{3n}\right)^{2/3}$$

(12)

The estimated values of $\mu$ and $m^*$ at 330 K are presented in **Table 2**. It is interesting to note that the best performing LSTO-H$_2$-in-C sample shows the highest effective mass ($m^*/m_0$ = 4.2) and highest drift mobility ($\mu$ = 10.9) among the three measured samples. These values are comparable to literature values reported for mobility ($\mu$ = 9.2) and effective mass ($m^*/m_0$ = 6.0) for a single crystal with a similar carrier concentration ($n$ = 6.8 × 10$^{20}$ cm$^{-3}$) at room temperature.[32]

The carrier transport and modelling results strongly indicate a reduction of the height of potential barriers at grain boundaries. To verify this reduction of the potential barrier, we used Kelvin probe force microscopy (KPFM), also known as scanning surface potential microscopy (SSPM) to characterize the contact potential difference (CPD) across the grain boundaries. The atomic force microscopy (AFM) height images of the sample surface were collected simultaneously during the KPFM measurement. **Figure 7** presents the height and CPD images collected from the LSTO-H$_2$ (**Figure 7**a and b) and LSTO-H$_2$-in-C (**Figure 7**d and e) samples. The grain boundaries are clearly distinguishable in both the height and CPD images due to the height variation of adjacent grains and the existence of potential barriers at the grain boundaries, respectively. The height profiles (**Figure 7**c and f) extracted from the lines marked in **Figure 7**a and d indicates a well-defined stair feature, showing a height variation around 2 nm between the adjacent grains. In contrast, the profiles of CPD do not follow that of the topography but exhibit a valley at the grain boundary region, suggesting negatively charged grain boundaries as expected.

More importantly, the magnitude of this negatively charged potential barrier at the grain boundary clearly differs between the two measured samples, ~ 6 mV for the LSTO-H$_2$ sample, and < 3 mV for the LSTO-H$_2$-in-C. The smaller magnitude of negative potential at the grain boundary of LSTO-H$_2$-in-C compared to that of the LSTO-H$_2$ sample correlates well with differences in their temperature dependency of conductivity and the $\Delta E_0$ values from the two-phase model (**Figure 6b**). Obviously, the measurement of surface potential by KPFM provides direct evidence that the LSTO-H$_2$-in-C sample with a single-crystal like carrier transport behaviour has a smaller potential barrier at the grain boundaries than the LSTO-H$_2$ with thermally activated conductivity. Nevertheless, there is a discrepancy between specific values of the measured potential barrier (several mV in scale) by KPFM and the band offset $\Delta E$ (10 to 100 meV) from the fitting of electrical conductivity with the two-phase model. This discrepancy is probably due to charge accumulation at the sample surface and the resolution limit of KPFM, causing the lowering and widening of potential profile.[46]

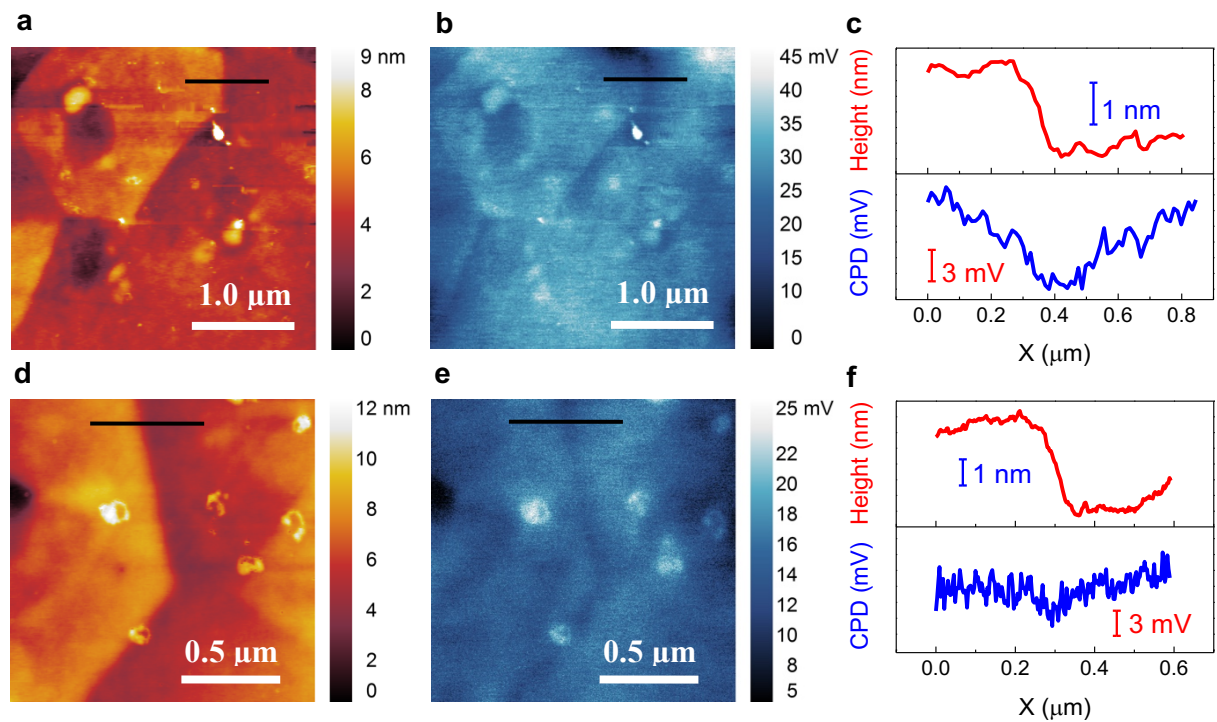

**Figure 7**. AFM and KPFM characterization for the LSTO-H$_2$ and LSTO-H$_2$-in-C samples. (a) AFM height and (b) the corresponding CPD images for the LSTO-H$_2$ sample with high grain boundary potential; (c) the extracted height (upper panel) and CPD (lower panel) profiles for the black lines marked in (a) and (b), respectively. (d) AFM height and (e) the corresponding CPD images for the LSTO-H$_2$-in-C sample with low grain boundary potential; (f) the extracted height (upper panel) and CPD (lower panel) profiles for the black lines marked in (d) and (e), respectively.

The two-phase model successfully reproduced the temperature dependent electrical conductivity of the LSTO samples with the $\Delta E_0$ defining the magnitude of band offset as the only fitting parameter. The measurement of grain boundary potential by KPFM confirmed the soundness of using the magnitude of band offset (i.e. potential barrier height) at grain boundary as a fitting parameter. We further simulated plots of (i) log |$S$|−log $\sigma$ and (ii) power factor as a function of log $\sigma$ at a temperature of 330 K, using the parameters listed in **Table 1**. The simulated plots are presented as lines in **Figure 8**; the symbols are experimental values from the three LSTO samples prepared in this work. The simulated plots fit well with the experimental results. According to **Figure 8**, it is clear that we achieved an order of magnitude enhancement in electrical conductivity with only minor sacrifice of Seebeck coefficients by the modulation of grain boundary properties. This conductivity enhancement further leads to an order of magnitude increase in power factor. Nevertheless, the power factor of the LSTO ceramic has not yet fully approached that of the single crystal. On the basis of the two-phase model, further enhancement of power factor is achievable by reduction of the size fraction of grain boundary phase $t_{GB}$ and/or the increase of the transport coefficient of grain boundary phase $\sigma_{E_0,GB}$.

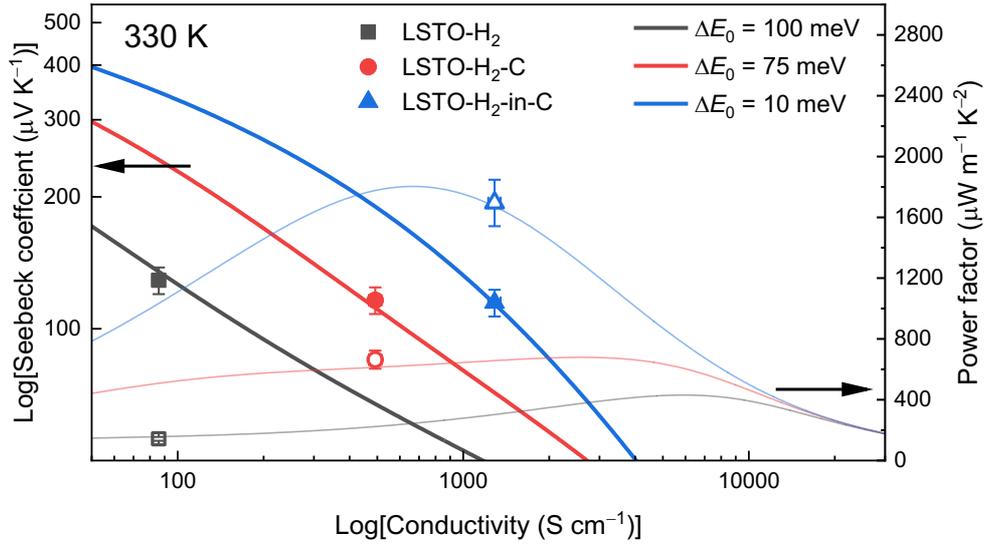

**Figure 8**. Modelled plots of (i) log Seebeck coefficients as a function of log electrical conductivity (thick solid lines) and (ii) power factor as a function of log electrical conductivity (feint lines) at room temperature (330 K) for the three LSTO ceramic samples prepared in the present work. The symbols are experimental results of log |S|—log σ (solid symbols) and power factor versus log σ (open symbols).

According to this significant enhancement of power factor at room temperature, the *ZT* value of LSTO-H$_2$-in-C increases by an order of magnitude to 0.07 from that of LSTO-H$_2$ (0.007) at 330 K (**Figure S6**). To achieve a high *ZT* value, it is necessary to reduce/maintain thermal conductivity of LSTO at a relatively low level while enhancing its electrical conductivity. Ideally, the lattice contribution to the thermal conductivity should remain unchanged when the electronic transport behaviour switches from a thermally activated to a single crystal-like behaviour. From the Wiedemann–Franz law, we estimated the electronic contribution ($\kappa_{electronic}$) of the thermal conductivity by $\kappa_{electronic} = L\sigma T$, where *L* is the Lorenz number and is determined by the transport equation (single parabolic band, acoustic phonon scattering *s* = 1):

$$L = \left(\frac{k_B}{e}\right) \frac{s(s+2)F_{s-1}(\eta)F_{s+1}(\eta) - (s+1)^2 F_s^2(\eta_i)}{s^2 F_{s-1}^2(\eta_i)}$$

(13)

The reduced Fermi energy level $\eta$ at each temperature was extracted from measured Seebeck coefficients according to equation (4). The lattice contribution ($\kappa_{lattice}$) to the total thermal conductivity ($\kappa_{total}$) is obtained by $\kappa_{lattice} = \kappa_{total} - \kappa_{electronic}$.

This deconvolution of electronic and lattice contributions to the total thermal conductivity (**Figure S7**a and b) indicates the higher total thermal conductivity of LSTO-H$_2$-in-C (8.1 W m$^{-1}$ K$^{-1}$) compared to that of LSTO-H$_2$ (6.9 W m$^{-1}$ K$^{-1}$) at 323 K is primarily due to the electronic contribution (0.8 W m$^{-1}$ K$^{-1}$). This much reduced impact of grain boundary modulation on the lattice thermal conductivity compared to electrical conductivity implies an opportunity to achieve the ultimate "phonon glass electron crystal" target by grain size refinement. Earlier work demonstrated a room temperature thermal conductivity as low as 1.2 W m$^{-1}$ K$^{-1}$ for La-doped (10 at.%) SrTiO$_3$ ceramic with a grain size of 24 nm.[9] Moreover, for nanostructured materials with grain size < 50 nm, energy filtering effects becomes prominent.[34-35] Earlier computational study indicates the opportunity to enhance both Seebeck coefficient and power factor via energy filtering effect by a precise control of both barrier height and grain size; and the optimum barrier height being around $k_BT$, ~26 meV at room temperature,[35] which is close to the modelled value in this work for LSTO-H$_2$-in-C sample (10 meV). However, the creation of a sufficiently high concentration of oxygen vacancies at the grain boundaries requires high temperature and long term annealing of LSTO in low oxygen partial pressure environments, which is in conflict with the need to limit grain growth. This challenging target might be achievable via the advancement of ceramic processing technologies and/or development of composites in which the secondary components (e.g. graphene) restrains grain growth while facilitating the creation of oxygen vacancies at the grain boundary.[6-7, 22, 40, 43]

**Conclusions**

We have shown that by both experimental results and modelling, the modulation of the built-in electrostatic potential at the grain boundaries of SrTiO$_3$ leads to charge carrier transport approaching that of single crystals. The two-phase model successfully accounted for the grain boundary potential as a band offset from the neutral grain, highlighting its dominant role in tuning the low temperature dependency of electrical conductivity. The model showed good agreement with the properties of experimentally prepared LSTO samples that exhibit gradual changes in electrical conductivity behaviour from thermally activated to single crystal type. KPFM imaging of the surface potential further confirmed the reduced grain boundary potential for LSTO samples with single crystal-like carrier transport behaviour. The successful modulation of grain boundary charge transport leads to an order of magnitude enhancement in the power factor value from 143 to 1745 µW m$^{-1}$ K$^{-2}$ at 330 K. This work provided further insights and understanding of grain boundary effects on the electrical charge transport in polycrystalline SrTiO$_3$ perovskite, and guidance in the design of effective routes to further enhance performance via grain boundary engineering.

ASSOCIATED CONTENT

Supporting Information

The supporting information is available free of charge on the ACS publications website at DOI:

Additional modelling details for Seebeck coefficients in series circuit model; computation of porosity corrected, effective electrical conductivity; energy band diagrams of two grains and their boundary region before and after contact; fitting of the literature reported single crystal carrier transport data with the carrier transport equations; plot of total Seebeck coefficient as a function of $t_{GB}$ in series circuit model; decoupled electric conductivity of neutral grain phase as a function of temperature for the LSTO samples; weighted mobility of literature reported La- or Nb- doped single crystals; dimensionless figure of merit for the LSTO samples; total, lattice and thermal conductivity for the LSTO samples.

AUTHOR INFORMATION

**Corresponding Authors**

*Email: ian.kinloch@manchester.ac.uk; robert.freer@manchester.ac.uk

**Author Contributions**

J.C. and D.E. contributed equally to this work. The manuscript was written through contributions of all authors. All authors have given approval to the final version of the manuscript.

**Notes**

The authors declare no competing financial interests.

ACKNOWLEDGEMENT

The authors are grateful to the EPSRC for the provision of funding for this work (EP/H043462, EP/I036230/1, EP/L014068/1, EP/L017695/1 acknowledged by RF). JC and IAK acknowledge the European Union's Horizon 2020 research and innovation program under grant agreements no. 785219 and 881603. IAK acknowledges Morgan Advanced Materials and the Royal Academy of Engineering for funding his chair. The work was also supported by the Henry Royce Institute for Advanced Materials, funded through EPSRC grants EP/R00661X/1, EP/S019367/1, EP/P025021/1 and EP/P025498/1. We gratefully acknowledge the support from X-ray facilities in Department of Materials in the University of Manchester. All research data supporting this work are directly available within this publication.

REFERENCES

(1) Kuo, J. J.; Kang, S. D.; Imasato, K.; Tamaki, H.; Ohno, S.; Kanno, T.; Snyder, G. J. Grain Boundary Dominated Charge Transport in $Mg_3Sb_2$-Based Compounds. *Energy Environ. Sci.* **2018,** *11*, 429-434.

(2) Xu, X.; Liu, Y.; Wang, J.; Isheim, D.; Dravid, V. P.; Phatak, C.; Haile, S. M. Variability and Origins of Grain Boundary Electric Potential Detected by Electron Holography and Atom-Probe Tomography. *Nat. Mater.* **2020,** *19*, 887-893.

(3) Dylla, M. T.; Kuo, J. J.; Witting, I.; Snyder, G. J. Grain Boundary Engineering Nanostructured $SrTiO_3$ for Thermoelectric Applications. *Adv. Mater. Interfaces* **2019,** *6*, 1900222.

(4) Raghuwanshi, M.; Wuerz, R.; Cojocaru-Mirédin, O. Interconnection between Trait, Structure, and Composition of Grain Boundaries in $Cu(In,Ga)Se_2$ Thin-Film Solar Cells. *Adv. Funct. Mater.* **2020,** *30*, 2001046.

(5) Youssef, K. M.; Scattergood, R. O.; Murty, K. L.; Koch, C. C. Ultratough Nanocrystalline Copper with a Narrow Grain Size Distribution. *Appl. Phys. Lett.* **2004,** *85*, 929-931.

(6) Rahman, J. U.; Du, N. V.; Nam, W. H.; Shin, W. H.; Lee, K. H.; Seo, W. S.; Kim, M. H.; Lee, S. Grain Boundary Interfaces Controlled by Reduced Graphene Oxide in Nonstoichiometric $SrTiO_{3-\delta}$ Thermoelectrics. *Sci. Rep.* **2019,** *9*, 8624.

(7) Lin, Y.; Dylla, M. T.; Kuo, J. J.; Male, J. P.; Kinloch, I. A.; Freer, R.; Snyder, G. J. Graphene/Strontium Titanate: Approaching Single Crystal–Like Charge Transport in Polycrystalline Oxide Perovskite Nanocomposites through Grain Boundary Engineering. *Adv. Funct. Mater.* **2020,** *30*, 1910079.

(8) Avila-Paredes, H. J.; Choi, K.; Chen, C.-T.; Kim, S. Dopant-Concentration Dependence of Grain-Boundary Conductivity in Ceria: A Space-Charge Analysis. *J. Mater. Chem.* **2009,** *19*, 4837-4842.

(9) Buscaglia, M. T.; Maglia, F.; Anselmi-Tamburini, U.; Marré, D.; Pallecchi, I.; Ianculescu, A.; Canu, G.; Viviani, M.; Fabrizio, M.; Buscaglia, V. Effect of Nanostructure on the Thermal Conductivity of La-Doped SrTiO$_3$ Ceramics. *J. Eur. Ceram. Soc.* **2014,** *34*, 307-316.

(10) Bell, L. E. Cooling, Heating, Generating Power, and Recovering Waste Heat with Thermoelectric Systems. *Science* **2008,** *321*, 1457-1461.

(11) Snyder, G. J.; Toberer, E. S. Complex Thermoelectric Materials. *Nat. Mater.* **2008,** *7*, 105-114.

(12) Koumoto, K.; Terasaki, I.; Funahashi, R. Complex Oxide Materials for Potential Thermoelectric Applications. *MRS Bull.* **2011,** *31*, 206-210.

(13) Heremans, J. P.; Dresselhaus, M. S.; Bell, L. E.; Morelli, D. T. When Thermoelectrics Reached the Nanoscale. *Nat. Nanotechnol.* **2013,** *8*, 471-473.

(14) Mori, T.; Priya, S. Materials for Energy Harvesting: At the Forefront of a New Wave. *MRS Bull.* **2018,** *43*, 176-180.

(15) Petsagkourakis, I.; Tybrandt, K.; Crispin, X.; Ohkubo, I.; Satoh, N.; Mori, T. Thermoelectric Materials and Applications for Energy Harvesting Power Generation. *Sci. Technol. Adv. Mater.* **2018,** *19*, 836-862.

(16) Soleimani, Z.; Zoras, S.; Ceranic, B.; Shahzad, S.; Cui, Y. A Review on Recent Developments of Thermoelectric Materials for Room-Temperature Applications. *Sustain. Energy Technol. Assess.* **2020,** *37*, 100604.

(17) Moos, R.; Härdtl, K. H. Electronic Transport Properties of $Sr_{1-x}La_xTiO_3$ Ceramics. *J. Appl. Phys.* **1996,** *80*, 393-400.

(18) Boston, R.; Schmidt, W. L.; Lewin, G. D.; Iyasara, A. C.; Lu, Z.; Zhang, H.; Sinclair, D. C.; Reaney, I. M. Protocols for the Fabrication, Characterization, and Optimization of N-Type Thermoelectric Ceramic Oxides. *Chem. Mater.* **2017,** *29*, 265-280.

(19) Kuo, J. J.; Yu, Y.; Kang, S. D.; Cojocaru-Mirédin, O.; Wuttig, M.; Snyder, G. J. Mg Deficiency in Grain Boundaries of n-Type $Mg_3Sb_2$ Identified by Atom Probe Tomography. *Adv. Mater. Interfaces* **2019,** *6*, 1900429.

(20) Wang, J.; Zhang, B. Y.; Kang, H. J.; Li, Y.; Yaer, X.; Li, J. F.; Tan, Q.; Zhang, S.; Fan, G. H.; Liu, C. Y.; Miao, L.; Nan, D.; Wang, T. M.; Zhao, L. D. Record High Thermoelectric Performance in Bulk $SrTiO_3$ Via Nano-Scale Modulation Doping. *Nano Energy* **2017,** *35*, 387-395.

(21) Li, J. B.; Wang, J.; Li, J. F.; Li, Y.; Yang, H.; Yu, H. Y.; Ma, X. B.; Yaer, X.; Liu, L.; Miao, L. Broadening the Temperature Range for High Thermoelectric Performance of Bulk Polycrystalline Strontium Titanate by Controlling the Electronic Transport Properties. *J. Mater. Chem. C* **2018,** *6*, 7594-7603.

(22) Lin, Y.; Norman, C.; Srivastava, D.; Azough, F.; Wang, L.; Robbins, M.; Simpson, K.; Freer, R.; Kinloch, I. A. Thermoelectric Power Generation from Lanthanum Strontium Titanium Oxide at Room Temperature through the Addition of Graphene. *ACS Appl. Mater. Interfaces* **2015,** *7*, 15898-15908.

(23) Okhay, O.; Zlotnik, S.; Xie, W.; Orlinski, K.; Hortiguela Gallo, M. J.; Otero-Irurueta, G.; Fernandes, A. J. S.; Pawlak, D. A.; Weidenkaff, A.; Tkach, A. Thermoelectric
32

Performance of Nb-Doped SrTiO$_3$ Enhanced by Reduced Graphene Oxide and Sr Deficiency Cooperation. *Carbon* **2019,** *143*, 215-222.

(24) Park, K.; Son, J. S.; Woo, S. I.; Shin, K.; Oh, M.-W.; Park, S.-D.; Hyeon, T. Colloidal Synthesis and Thermoelectric Properties of La-Doped SrTiO$_3$ Nanoparticles. *J. Mater. Chem. A* **2014,** *2*, 4217-4224.

(25) Mendelson, M. I. Average Grain Size in Polycrystalline Ceramics. *J. Am. Ceram. Soc.* **1969,** *52*, 443-446.

(26) Sharma, P. A.; Brown-Shaklee, H. J.; Ihlefeld, J. F. Oxygen Partial Pressure Dependence of Thermoelectric Power Factor in Polycrystalline N-Type SrTiO$_3$: Consequences for Long Term Stability in Thermoelectric Oxides. *Appl. Phys. Lett.* **2017,** *110*, 173901.

(27) Ekren, D.; Azough, F.; Gholinia, A.; Day, S. J.; Hernandez-Maldonado, D.; Kepaptsoglou, D. M.; Ramasse, Q. M.; Freer, R. Enhancing the Thermoelectric Power Factor of Sr$_{0.9}$Nd$_{0.1}$TiO$_3$ through Control of the Nanostructure and Microstructure. *J. Mater. Chem. A* **2018,** *6*, 24928-24939.

(28) Marrocchelli, D.; Sun, L.; Yildiz, B. Dislocations in SrTiO$_3$: Easy to Reduce but Not So Fast for Oxygen Transport. *J. Am. Chem. Soc.* **2015,** *137*, 4735-4748.

(29) Pike, G. E.; Seager, C. H. The Dc Voltage Dependence of Semiconductor Grain-Boundary Resistance. *J. Appl. Phys.* **1979,** *50*, 3414-3422.

(30) Kang, S. D.; Snyder, G. J. Charge-Transport Model for Conducting Polymers. *Nat Mater* **2017,** *16*, 252-257.

(31) Moos, R.; Menesklou, W.; Härdtl, K. H. Hall Mobility of Undoped N-Type Conducting Strontium Titanate Single Crystals between 19 K and 1373 K. *Appl. Phys. A* **1995,** *61*, 389-395.

(32) Ohta, S.; Nomura, T.; Ohta, H.; Koumoto, K. High-Temperature Carrier Transport and Thermoelectric Properties of Heavily La- or Nb-Doped $SrTiO_3$ Single Crystals. *J. Appl. Phys.* **2005,** *97*, 034106.

(33) Kim, S.; Fleig, J.; Maier, J. Space Charge Conduction: Simple Analytical Solutions for Ionic and Mixed Conductors and Application to Nanocrystalline Ceria. *Phys. Chem. Chem. Phys.* **2003,** *5*, 2268-2273.

(34) Thesberg, M.; Kosina, H.; Neophytou, N. On the Effectiveness of the Thermoelectric Energy Filtering Mechanism in Low-Dimensional Superlattices and Nano-Composites. *J. Appl. Phys.* **2016,** *120*, 234302.

(35) Kim, R.; Lundstrom, M. S. Computational Study of Energy Filtering Effects in One-Dimensional Composite Nano-Structures. *J. Appl. Phys.* **2012,** *111*, 024508.

(36) Lu, Z.; Zhang, H.; Lei, W.; Sinclair, D. C.; Reaney, I. M. High-Figure-of-Merit Thermoelectric La-Doped A-Site-Deficient $SrTiO_3$ Ceramics. *Chem. Mater.* **2016,** *28*, 925-935.

(37) Waser, R. Electronic Properties of Grain Boundaries in $SrTiO_3$ and $BaTiO_3$ Ceramics. *Solid State Ion.* **1995,** *75*, 89-99.

(38) Yaremchenko, A. A.; Populoh, S.; Patrício, S. G.; Macías, J.; Thiel, P.; Fagg, D. P.; Weidenkaff, A.; Frade, J. R.; Kovalevsky, A. V. Boosting Thermoelectric Performance by Controlled Defect Chemistry Engineering in Ta-Substituted Strontium Titanate. *Chem. Mater.* **2015,** *27*, 4995-5006.

(39) Zhang, L.; Tosho, T.; Okinaka, N.; Akiyama, T. Thermoelectric Properties of Solution Combustion Synthesized Al-Doped ZnO. *Mater. Trans.* **2008,** *49*, 2868-2874.

(40) Feng, X.; Fan, Y.; Nomura, N.; Kikuchi, K.; Wang, L.; Jiang, W.; Kawasaki, A. Graphene Promoted Oxygen Vacancies in Perovskite for Enhanced Thermoelectric Properties. *Carbon* **2017,** *112*, 169-176.

(41) Kovalevsky, A. V.; Aguirre, M. H.; Populoh, S.; Patrício, S. G.; Ferreira, N. M.; Mikhalev, S. M.; Fagg, D. P.; Weidenkaff, A.; Frade, J. R. Designing Strontium Titanate-Based Thermoelectrics: Insight into Defect Chemistry Mechanisms. *J. Mater. Chem. A* **2017,** *5*, 3909-3922.

(42) Azough, F.; Gholinia, A.; Alvarez-Ruiz, D. T.; Duran, E.; Kepaptsoglou, D. M.; Eggeman, A. S.; Ramasse, Q. M.; Freer, R. Self-Nanostructuring in $SrTiO_3$: A Novel Strategy for Enhancement of Thermoelectric Response in Oxides. *ACS Appl. Mater. Interfaces* **2019,** *11*, 32833-32843.

(43) Wu, C.; Li, J.; Fan, Y.; Xing, J.; Gu, H.; Zhou, Z.; Lu, X.; Zhang, Q.; Wang, L.; Jiang, W. The Effect of Reduced Graphene Oxide on Microstructure and Thermoelectric Properties of Nb-Doped a-Site-Deficient $SrTiO_3$ Ceramics. *J. Alloys Compd.* **2019,** *786*, 884-893.

(44) Teranishi, T.; Ishikawa, Y.; Hayashi, H.; Kishimoto, A.; Katayama, M.; Inada, Y. Thermoelectric Efficiency of Reduced $SrTiO_3$ Ceramics Modified with La and Nb. *J. Am. Ceram. Soc.* **2013,** *96*, 2852-2856.

(45) Snyder, G. J.; Snyder, A. H.; Wood, M.; Gurunathan, R.; Snyder, B. H.; Niu, C. Weighted Mobility. *Adv. Mater.* **2020,** *32*, 2001537.

(46) Kalinin, S. V.; Bonnell, D. A. Screening Phenomena on Oxide Surfaces and Its Implications for Local Electrostatic and Transport Measurements. *Nano Lett.* **2004,** *4*, 555-560.

**For Table of Contents Only (8.25 cm by 4.45 cm)**

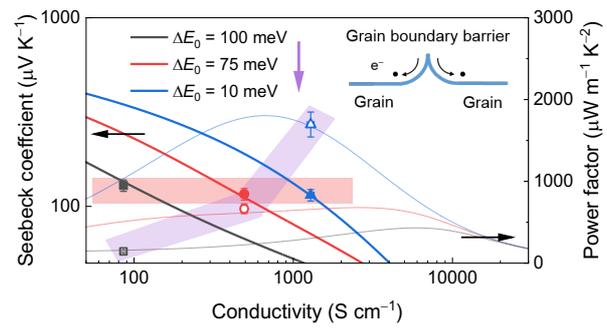